\documentclass[10pt,journal,compsoc]{IEEEtran}

\usepackage{amssymb}
\usepackage{amsmath}
\usepackage{array}
\usepackage{graphicx}
\usepackage{flushend}

\newcommand{\real}{\mathbb{R}}
\newcommand{\EXP}[1]{{E\left[{#1}\right]}}

\newcommand{\hw}{{\hat{\bw}}}

\newcommand{\id}{{\mathbf{I}}}

\newcommand{\linrel}{\mbox{\sc LinRel}}

\newcommand{\br}{{\mathbf{r}}}

\newcommand{\bK}{{\mathbf{K}}}
\newcommand{\bw}{{\mathbf{w}}}
\newcommand{\bweye}{{\mathbf{v}}}
\newcommand{\bwmkl}{{\mathbf{f}}}

\newcommand{\ba}{{\mathbf{a}}}
\newcommand{\be}{{\psi(I)}}

\newcommand{\etab}{\boldsymbol{\eta}}

\newcommand{\Alpha}{\boldsymbol{\alpha}}
\newcommand{\Beta}{\boldsymbol{\beta}}
\newcommand{\ld}{{\lambda}}

\newcommand{\Kcal}{\mathcal{K}}
\newcommand{\ie}{{\em i.e.\/}}

\newcommand{\LP}{\left(}
\newcommand{\RP}{\right)}

\begin{document} 

\title{PinView: Implicit Feedback in Content-Based Image Retrieval}

\author{Zakria~Hussain$^1$,
        \and Arto~Klami$^{2,3}$,
        \and Jussi~Kujala$^2$,
        \and Alex~P.~Leung$^4$,
        \and Kitsuchart~Pasupa$^5$,
        \and Peter~Auer$^4$,
        \and Samuel~Kaski$^{2,3}$,
        \and Jorma~Laaksonen$^2$,
        \and John~Shawe-Taylor$^1$\thanks{The author order is first postdocs,
then PI's, both in alphabetical order.}
\thanks{$^1$University College London, United Kingdom;
$^2$Aalto University, Finland;
$^3$University of Helsinki, Finland;
$^4$University of Leoben, Austria;
$^5$King Mongkut's Institute of Technology Ladkrabang,
Thailand
}
}

\maketitle

\begin{abstract} 
This paper describes PinView, a content-based image retrieval system
that exploits implicit relevance feedback collected during a search
session.  PinView contains several novel methods to infer the intent
of the user.  From relevance feedback, such as eye movements or
pointer clicks, and visual features of images, PinView learns a
similarity metric between images which depends on the current
interests of the user.  It then retrieves images with a specialized
online learning algorithm that balances the tradeoff between exploring
new images and exploiting the already inferred interests of the user.
We have integrated PinView to the content-based image retrieval system
PicSOM, which enables applying PinView to real-world image databases.
With the new algorithms PinView outperforms the original PicSOM, and
in online experiments with real users the combination of implicit and
explicit feedback gives the best results.
\end{abstract} 


\section{Introduction}

Finding interesting images from a large collection is
a need common to both laymen surfing the internet and professional
graphic designers in their work. Content-based image retrieval (CBIR)
systems try to fulfil the need by showing images that are similar to
what they think the user is looking for.  Unfortunately CBIR systems typically
do not have a clear idea of what the user is looking for, due to
difficulties in communicating visual content from the user to the
system. One approach to get the search started is to prune the set of
candidate images with a keyword or tag search, given that the
images in the collection have been tagged beforehand.  However, simple
tags do not carry all interesting information about the images and
hence the number of images after pruning might still be so large that
a further content-based search is needed. Moreover, the tags are often
imperfect and it may be difficult for the user to formulate relevant
tags, for example if the goal is to find an image that fits well into
a given poster, or to find a beautiful image of a flower.  A common
approach to continue the search, after possible pruning by tags, is to
ask the user for explicit relevance feedback on the shown
images~\cite{datta:acm08}.  However, this is a laborious process and
the user might be unwilling to invest such an effort, for example
while casually browsing the web.

Another approach is to obtain this feedback implicitly, by measuring
indirect signals on attention patterns of the users and inferring the
relevance of the seen images from these~\cite{%
  DBLP:phd/de/Essig2007,GrecuPhD06,diane:sigir02,klami:mir08,OyekoyaPhD07,Scherffig05}.
This is the approach taken also by PinView, a CBIR system presented in
this paper. PinView uses implicit feedback from eye movements and
explicit feedback from pointer clicks to infer the interests of the
user, in order to iteratively show more relevant images.
Our argument on the feasibility of using eye tracking is that even
though trackers are still somewhat expensive and cumbersome to use, it
is a plausible scenario that they will become widely available and
widely used in several applications. There are no fundamental
restrictions on why an eye tracker could not be integrated into every
PC and smart phone, because mass-manufacturing core components of
trackers based on infrared oculography is not expensive: they require
a camera, an infra-red light source, and software. Once eye trackers
are available, the added cost of using them in CBIR is very low.

PinView must solve several subproblems to take advantage of the
recorded noisy implicit relevance feedback.  The first problem is how
to infer relevance of seen images from gaze patterns and clicks.  The
second problem is that there is a multitude of different visual (low-level)
features and similarity measures for images.  Each of these captures a
specific aspect of similarity, like color, texture, or shape of edges
in the image.  Which of these features are relevant in the current
search and how does the user perceive them?  PinView infers a
customized similarity metric for each search session with a multiple
kernel learning algorithm and tensor projection working on these
features.

The third and final problem is how to select images to show to the
user.  Given that the system is able to show the user only a
limited number of images, how should it balance exploitation of its
currently limited knowledge of the query and exploration of new kinds
of images?  PinView incorporates a specialized
exploration-exploitation algorithm \linrel\ which uses the inferred
metric between images to suggest new images to be shown to the user.

In this paper we introduce the full PinView system expanding from the
two partial views in preliminary conference papers
\cite{Auer10wapa,Hussain10}. We apply PinView to both offline and
online CBIR tasks to study it both in controllable setups and in real
image retrieval settings.

The structure of the remaining paper is as follows.  First, in
Section~\ref{section:background} we go through related work.
Section~\ref{section:system} presents details on the different
components of PinView.  Section~\ref{section:experiments} gives
results of the experiments with the PinView system and finally
Section~\ref{section:conclusion} concludes.

\section{Background and Related Work}
\label{section:background}

In this section we discuss related work in image retrieval and
eye movement research.
Content-based image retrieval (CBIR) is a well-researched topic, whose
history can be followed and comprehensive introductions to which can
be found in surveys such
as~\cite{datta:acm08,Rui99,Smeulders2000,Veltkamp2002}.
In addition, many of CBIR's research questions have been
covered by related works on content-based multimedia retrieval,
including, e.g., reviews~\cite{Lew2006,Sebe2003}.

The \emph{semantic gap}~\cite{Smeulders2000}, i.e., the unavoidable
inability of low-level visual features to capture and mediate the
semantic content and mutual similarity of images, has often been cited
as the foremost hindrance of successful CBIR.
In particular, the existence of the semantic gap has been given as a
reason for the failure of image retrieval approaches that have relied
on automatic image interpretation and textual querying.
How severely the gap actually harms the accuracy and usability of a
CBIR system will depend on the application and the particular image
retrieval task at hand.
In some types of searches it will be just the visual and not the
semantic similarity between the searched and retrieved images that
plays the primary role and, consequently, the problem of the semantic
gap will be minimal~\cite{datta:acm08}.

Since the mid-1990's, \emph{relevance feedback}
has been used for incorporating the user's preferences
and his understanding of the semantic similarity of images in the
retrieval process~\cite{Picard96-2,Rui98-2}.
Research on relevance feedback techniques constitutes a subfield of
CBIR research in its own right and the early works on the topic have
been summarized in~\cite{Zhou2003}.
The forms of explicit user interaction and giving of relevance
feedback in interactive CBIR vary.
In retrieval systems with multiple feature representations of the
images, a straightforward approach could be to ask the user to tune
the relative weights of the features in order to be able to find more
relevant images~\cite{Rui98-2}.
The weight tuning method and other approaches where the user is
required to be able to modify the internal parameters of the CBIR
system are, however, impractical for non-professional use.

In practical CBIR systems implementing relevance feedback, the
standard setting is that after the user has been presented with a set of
images, the system expects him to reliably assess the relevance of
each retrieved image and to return this information back to the
system~\cite{Zhou2003}.
This effectively reformulates the interactive image retrieval process
as an online machine learning task with small but increasing numbers
of training samples to learn the statistics of relevant (and
non-relevant) images on each query round.
From the user interface perspective, this type of relevance feedback
is often implemented by the means of the user clicking on the relevant
images, checking associated check boxes or giving a numerical relevance
assessment to each image with a slider or from a multi-value choice
list.
It is also possible that instead of assessing each image
independently, the user is asked to rank the images on the page
by their relevance in \emph{comparison searching}~\cite{Cox98}.

In numerous studies (e.g.~those cited in~\cite{Zhou2003}), explicit
interactive relevance feedback has been shown to provide a dramatic
improvement in the accuracy of image retrieval.
Giving explicit and accurate relevance feedback for each seen image
is, however, bound to be time consuming and cognitively strenuous.
Therefore, implicit feedback strategies have received considerable
interest in the information retrieval (IR) community, due to the
promise of decreasing the burden on the user. It has become clear that
implicit feedback can improve information retrieval accuracy (see the
review~\cite{diane:sigir02}), but figuring out the most effective
modalities for various search scenarios is still a subject of ongoing
research and various alternatives are being proposed ranging from
simple measures like number of clicks to brain computer interfaces
that are not yet practically feasible for real search tools.

The more traditional implicit feedback approaches rely on feedback
obtained from the control devices. Claypool et al.~\cite{Claypool01}
studied use of mouse and keyboard activity, as well as time spent on
the page and scrolling, and \cite{Fox05} compared the amount of
information between such implicit channels and explicit feedback. The
most consistent finding in these kinds of works has been that the time
spent on the page and the way the user exits the page are good
indicators of relevance. More advanced works still using the regular
control devices use click-through data, typically on the search result
page \cite{Joachims05}. While these sources of implicit information
are readily available for all search tools, they provide a rather
limited view of the actions and intents of the user.

In the other extreme, a number of approaches have used brain computer
interfaces for IR or related tasks. The C3Vision system
\cite{Gerson06} and a human-aided computing approach by
\cite{Shenoy08} infer image categories or presence of distinct objects
in images from EEG measurements, and \cite{Kay08,Mitchell04} use fMRI
techniques for image categorization. Wang et al.~\cite{Wang09} built
a prototype image annotation system using these ideas; relevance of
images is inferred from EEG and visual pattern mining is used to
retrieve similar images.  They do not, however, consider a full
relevance feedback procedure for retrieval, but only study a single
iteration and measure the performance as annotation accuracy. Brain
activity measurements provide the most accurate picture of the intents
of the user, but are clearly not yet practically feasible for real
retrieval tools. Notable instrumentation and modeling challenges
remain to be solved for making the devices applicable for daily use.

The most interesting implicit feedback modalities fall between these
two extremes. Various information signals can be captured by
microphones, cameras or other easily wearable sensors, and they are
likely to contain more information on the intentions of the user than
what can be observed through the traditional control devices. Both
speech and gestures have been extensively used as explicit control
modalities, but there are also a few studies on their implicit use.
For example, \cite{Vinciarelli09} infers tags for images from implicit
speech and \cite{Arapakis09} considers facial expressions as
indicators of topical relevance. In addition, various physiological
measurements are extensively used for inferring the affective state of
the user, which can in turn be used as a feedback source
\cite{Arapakis08,Soleymani08}. However, to our knowledge there are no
fully fledged image retrieval systems that use these input
modalities as implicit feedback.

The primary feedback in this work is based on eye movements, which
have become an increasingly popular feedback source in recent years,
following the early concepts by \cite{Maglio00}. The primary body of
eye-tracking works in IR has been done for text retrieval,
because the highly structured eye movements while reading are easier
to model. The
approaches range from explicit control \cite{Ward02} and
relevance estimation of text passages \cite{Buscher08,Puolamaki05sigir}
to inferring complete queries based on eye-movements on the results
pages \cite{Ajanki09}.  

The text retrieval works were followed by early attempts to utilizing
eye movements in image retrieval.
Based on the results of a comparison between a visual attention model
and measured gaze fixations, it was suggested in~\cite{Oyekoya2004}
that eye tracking could be used as an interface for image retrieval,
but no actual retrieval setup was yet investigated.
The \emph{Eye-Vision-Bot} system, presented in~\cite{Scherffig05},
integrated an eye tracker with the GIFT image retrieval
system\footnote{http://www.gnu.org/software/gift/} merely as a
demonstration of the possibilities of gaze-based interaction without
any experimental evaluations.
In~\cite{GrecuPhD06,Grecu05}, a CBIR system was implemented that used
offline image saliency and online gaze fixations for extracting visual
features from those image areas that were likely to be relevant when
determining the relevancy of the image.  The system showed promising
results in offline experiments, but was not ready for real interactive
user experiments.

First fully interactive and experimentally evaluated CBIR systems that
made use of eye-tracking data were presented in~\cite{%
  DBLP:phd/de/Essig2007,OyekoyaPhD07,Oyekoya06}.
The selection of an image as relevant was in~\cite{Oyekoya06} solely
dependent on the accumulated fixation time exceeding a preset
threshold, whereas in~\cite{OyekoyaPhD07} also a richer set of gaze
parameters, including saccadic speeds and the number of images with
fixations, were used.
Image similarity assessment was in~\cite{DBLP:phd/de/Essig2007} based
on visual features extracted from non-overlapping tiles of the images.
The user indicated the most relevant image by clicking, after which
new images were retrieved based on the sum of tile-wise feature
distances weighted with values from the fixation map.
Clear performance improvements were obtained in the evaluations over
random selection in~\cite{OyekoyaPhD07,Oyekoya06} and over simple
image clicking without gaze-based distance weighting
in~\cite{DBLP:phd/de/Essig2007}.

Two decisive characteristics common to the setups of~\cite{%
  DBLP:phd/de/Essig2007,OyekoyaPhD07,Oyekoya06} should, however, be
noticed.  First, the user is expected to always explicitly select
exactly one relevant image, by either eye fixation or mouse
clicking. Second, the user interface has in the experiments been such
that the target or query image is continuously visible on the screen,
which is not plausible in real CBIR applications.  Showing the target
will also facilitate and even encourage the use of gaze for image
comparison, which will certainly have an effect on the gaze patterns.

Later, also~\cite{Kozma09} and~\cite{Liang10,Zhang10} and~\cite{Faro10} 
introduced their image retrieval systems using eye
movements. 
The first one~\cite{Kozma09} was based on a conceptual interface
designed to be controlled completely by implicit gaze, providing a mix
of a browsing and search tool. A small-scale online experiment was
provided, but it cannot be used for drawing strong conclusions on the
accuracy of the retrieval results.
The second study mostly concentrated on the accuracy of inferring the
relevance in~\cite{Zhang10} and on fixation-weighted region matching
between the query and database images in~\cite{Liang10}.
The last one~\cite{Faro10} used gaze data for genuinely implicit
relevance feedback by the means of reranking the results of Google
Image Search.  However, the system was not fully functional yet as the
described experimantal evaluation was done in a non-interactive mode.

\section{System Components} 
\label{section:system}

\begin{figure*}[ht]
\begin{center}
\centerline{\includegraphics[width=1.5\columnwidth]{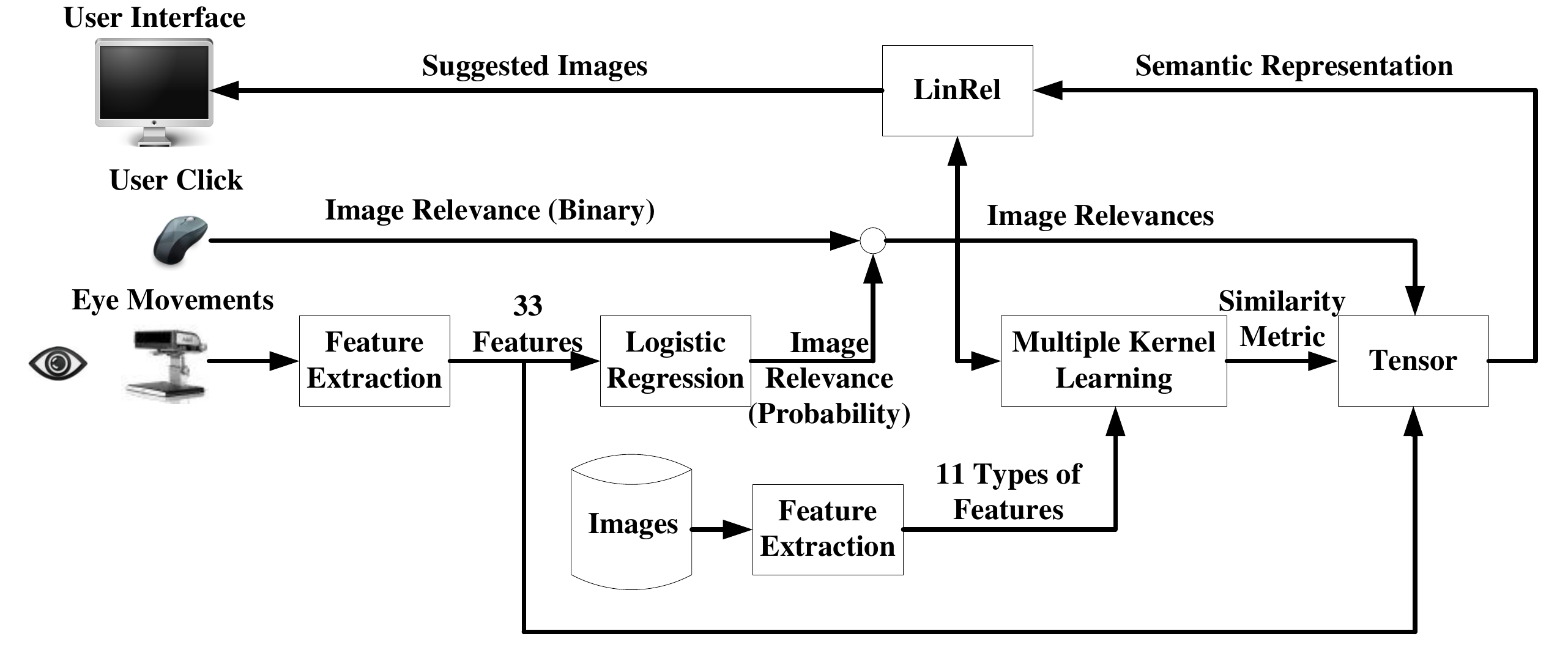}}
\caption{Main components and data flow in PinView.}
\label{figure:relations}
\end{center}
\vskip -0.2in
\end{figure*} 

In this section we describe the main components of the system.
It consists of four main components, which
will be explained in more detail in the following sections.
The first component predicts the relevance of seen images
based on clicks and image features.
Tensor decomposition and multiple kernel learning modules then
infer a metric between images using known visual features of the
images (see Table~\ref{table:imagefeatures} and~\cite{SPIC2007} for
more detailed descriptions of the used features) and relevance
feedback on the seen images. The final component, a specialized
exploration-exploitation algorithm \linrel\ suggests new images to be
shown to the user.

Figure~\ref{figure:relations} summarizes the flow of information and
the relationships between the different components. The input from the
user, captured by mouse clicks and the eye tracker, is fed into the
image relevance predictor. The predicted relevance scores are then
given to the multiple kernel learning module together with the image
features extracted from the images, for the purpose of learning which
feature sets the similarity metric should utilize for comparing the
images. The metric is fed to the tensor decomposition module to be
combined with the eye movement features, in order to learn a
representation that enables implicitly estimating eye movement
features also for unseen images. Finally, the system selects a
new set of images with the \linrel\ algorithm based on the inferred
relevance scores and the final metric given by the tensor
decomposition, and the images are retrieved from a database and
displayed through the PicSOM backend~\cite{laaksonen:networks02}.

\begin{table}
\caption{Image features used in PinView}
\label{table:imagefeatures}
\begin{center}
\begin{tabular}{| l |r|} \cline{1-2}
\multicolumn{1}{| l |}{Feature} & \multicolumn{1}{c|}{dim.} \\\cline{1-2}
DCT coefficients of average colour in rectangular grid & 12  \\
CIE Lab colour of two dominant colour clusters & 6 \\
Histogram of local edge statistics & 80 \\
Haar transform of quantised HSV colour histogram & 256 \\
Histogram of interest point SIFT features & 256 \\
Average CIE Lab colour & 15 \\
Three central moments of CIE Lab colour distribution & 45 \\
Histogram of four Sobel edge directions & 20 \\
Co-occurrence matrix of four Sobel edge directions & 80 \\
Magnitude of the $16 \times 16$ FFT of Sobel edge image & 128 \\
Histogram of relative brightness of neighbouring pixels & 40 \\
\cline{1-2}
\end{tabular}
\end{center}
\end{table}

\subsection{Relevance Prediction from Eye Movements and Clicks}
\label{section:relevance}

PinView infers relevance of images during a search task from implicit
feedback, explicit feedback given be the user, or their
combination. As implicit feedback PinView uses eye movements of the
user, building on the recent promising results on inferring image
relevance from eye movements \cite{Kozma09,Zhang10}. 

The gaze direction is an indicator of the focus of attention, since
accurate viewing is possible only in the central fovea area which
covers 1--2 degrees of the visual angle.
However, the correspondence is not one-to-one because the users can
shift the attention without moving their eyes.
Gaze tracking has been used extensively in the psychology literature,
and more recently also in information retrieval settings to track
attention patterns of users.
Some examples include the human-computer interaction aspects of how
users perform searches~\cite{cutrell07}, analysis of user behavior in
web search~\cite{granka04}, and using eye movements as implicit
relevance feedback in textual IR~\cite{Hardoon07,Salojarvi03}.
The promising results on the textual IR task suggest that using eye
movements for relevance determination could be possible also in image
retrieval tasks, where they would be even more severely needed.
Hence, the PinView system estimates the relationship between eye
movement patterns and relevance of images from data.  As explicit
feedback PinView uses pointer clicks by the user.

We measured the eye movements with a Tobii 1750 eye tracker with 50Hz
sampling rate.  The tracker has two infra-red lights and an infra-red
stereo camera attached to a flat-screen monitor, and the
tracking is based on detection of pupil centers and measurement of
corneal reflection.  The eyes move in rapid ballistic movements called
saccades, from one fixation to another. Within each fixation the eyes
are fairly motionless.  Raw eye measurements are preprocessed by first
extracting fixations and saccades, judging a set of consecutive raw
measurements to be a fixation if they occur within a dispersion
of 30 pixels, which at normal viewing distance is equivalent to
roughly 0.6 visual degrees (17 inch screen with resolution of
1280$\times$1024 pixels).  A fixation is defined to be a period of at
least 100 milliseconds of looking at a single location on the screen.

Inferring the relevance feedback requires a mapping from the gaze
pattern to the relevances. It is infeasible to assume that such a
mapping could be constructed from first principles of human vision,
and therefore we take the machine learning approach of learning it
from data. That is, we assume a simple parametric mapping from a set
of gaze features computed from the eye movement trajectory to the
relevances, and learn its parameters from a training
data with known relevance scores.
To avoid needing user adaptation, we learned a single user- and
task-independent predictor from data collected from multiple users and
a few search tasks. This was done on data collected in online search
sessions separate from the actual experiments reported in
Section~\ref{section:experiments}, to avoid possible biases due to
having trained the relevance predictor in the same search tasks.

For each viewed image $I$ PinView collects 19 features
(Table~\ref{table:eyefeatures}) computed from both raw eye movement
samples and fixations, including aspects such as the logarithm of the
total time the image was looked at and the number of regressions to
already seen images. Instead of attempting to construct maximally
pyschovisually motivated features, the set of features was chosen so
that they are efficient to compute and can intuitively be expected to
be informative of the relevance. Furthermore, the features do not
depend on the image content, so that the predictor can directly
generalize to different search tasks and databases.

The relevance score of an image is predicted from the
features using a logistic regression model.
In detail, for an eye movement feature vector $\be$ PinView computes
the relevance score $\textrm{rel}(\be|\bweye,b)$ as
\[
\textrm{rel}(\be|\bweye,b)  = \frac{1}{1+\exp{\left(-\bweye^\top \be +b\right)}},
\]
where $\bweye$ and $b$ are the learned parameters, a weight
vector and a bias term. 
To improve the accuracy, each feature was standardized to zero mean
and unit variance and the parameters were learned with 2-norm
regularization on the weights, the regularization constant selected
by 5-fold cross validation.
Finally, the predicted relevance for images not viewed at all is
set at a small constant value.

In the relevance predictor training data, six subjects (staff members
of Aalto University who were not associated with this work) performed
12 different search tasks.  The objective of each task was to find as
many examples as possible of a given image category of the PASCAL
Visual Object Classes Challenge 2007 (VOC2007)
dataset~\cite{pascal-voc-2007}.  Ten collages consisting of 15 images
chosen by the PicSOM system were shown in each task, containing
a varying number of relevant images 
to cover various types of collages observed in real search tasks.
In six of the 12
search tasks the objective was to find either a cat or dog and the
database was limited to cat and dog images, resulting in around 50\%
of images being relevant. The other six tasks had 8--12\% of relevant
images, and the target was either motorbikes or aircrafts in the full
VOC 2007 collection.

Finally, when combining implicit and explicit feedback, we resorted to
a simple and fast method: The information of which images were clicked
is integrated to the model by adding a constant $\alpha$, determined
in offline experiments, to the relevance score of the clicked
image. The final relevance prediction is hence given by
\begin{equation}
\textrm{rel}(\be,\delta|\bweye,b,\alpha)  = \frac{1}{1+\exp{\left(-\bweye^\top \be +b\right)}} + \delta \alpha ,
\label{eq:eye-prediction}
\end{equation}
where $\delta=1$ for images that were clicked and $\delta=0$ for all other
images. As a side effect, the relevance score is not directly
interpretable as a probability but that does not affect the next steps.

\begin{table*}[t]
\caption{Eye movement features collected in PinView.}
\label{table:eyefeatures}
\begin{center}
\begin{tabular}{lll}
\hline
Number & Name & Description\\
\hline
\multicolumn{3}{l}{\textbf{Raw data features}} \\
1  & numMeasurements  & log of total time of viewing the image \\
2 & numOutsideFix & total time for measurements outside fixations \\
3 & ratioInsideOutside  & percentage of measurements inside/outside fixations \\
4 & speed & average distance between two consecutive measurements \\
5 & coverage & number of subimages ($4 \times 4$ grid) that contain at least one measurement\\
6 & normCoverage & coverage normalized by numMeasurements \\
7 & pupil & maximal pupil diameter during viewing \\
8 & nJumps1 & number of breaks (measurements outside the image between two visits) longer than 60ms \\
9 & nJumps2 & number of breaks longer than 600ms\\
\hline
\multicolumn{3}{l}{\textbf{Fixation features}} \\
10 & numFix & total number of fixations \\
11 & meanFixLen & mean length of fixations \\
12 & totalFixLen & total length of fixations\\
13 & fixPrct & percentage of time spent in fixations\\
14 & nJumpsFix & number of re-visits (regressions) to the image\\
15 & maxAngle & maximal angle between two consecutive saccades, transitions from one fixation to another \\
16 & firstFixLen & length of the first fixation\\
17 & firstFixNum & number of fixations during the first visit\\
18 & distPrev & distance to the fixation before the first visit\\
19 & durPrev & duration of the fixation before the first visit\\
\hline\\
\end{tabular}
\end{center}
\end{table*}

\subsection{Multiple Kernel Learning}
\label{section:mkl}

Learning the similarity measures or metric of importance for our CBIR
task is central in retrieval.  Some image searches may require a
combination of image features to quickly distinguish them from other
less relevant images.  For instance, colour and texture features may
be important to find pictures of snowscapes, whereas colour may be the
only important feature needed to find images of blue skies.  We would
like to use a combination of the metrics as a cue to finding relevant
images quickly and efficiently, and then pass this learnt metric
(kernel) to the \linrel~algorithm of Section~\ref{s:LINRELAlgorithm}.

Given image feature vectors $\phi(I)$, $\phi(J)$, let the inner
product $\phi(I) \phi(J)^\top = k(I,J)$ denote the kernel function
$k(\cdot,\cdot)$ between images $I$ and $J$, where $\phi$ is some
feature mapping~\cite{stc-04}. Multiple kernel learning (MKL) attempts
to find a combination of kernels by solving a classification (or
regression) problem using a weighted combination of
kernels~\cite{Argyriou-05,Bach-04,Lanckriet-04}.  Given that our
PinView system will use several different image features,
we view each one as a separate feature
space -- hence, giving us $N$ different kernels \ie, $\Kcal =
\{k_1,\ldots,k_{N}\}$, for the $N$ different image features.  Using MKL we construct the kernel function:
\[
k_{\etab}(I,J) = \sum_{i=1}^{N} \eta_i\, k_i(I,J),
\]
where $\etab = (\eta_1,\ldots,\eta_N)$ are the weights of each kernel
function $k_i(I,J)$ between images $I$ and $J$.  We follow an elastic-net
formulation of ridge regression MKL,
which uses a parameter $\lambda \in
[0,1]$ in order to move between a 1-norm regularization (when $\lambda
= 1$) and a 2-norm regularization (when $\lambda = 0$).
\footnote{We would be able to dynamically change the value of
  $\lambda$ throughout the search, however for simplicity we will fix
  $\lambda=0.5$ in the experiments.}  
Let $\Phi_t = [\phi(I_\tau)]_{\tau=1,\ldots,t-1}$ be the Gram matrix
of image features, and let $\br_t=(r_{I_1},\ldots,r_{I_{t-1}})$ be the
vector of relevance scores observed so far, then we solve the
following multiple kernel learning regression problem:
\begin{align}\label{mkl}
\min_{\bwmkl_i,\boldsymbol{\eta}} & \quad  \sum_{i=1}^{N} \LP \frac{\ld}{\eta_i} + 1 - \ld \RP \| \bwmkl_i \|^2_2 +  \Big\|  \sum_{i=1}^{N} \Phi_t^\top \bwmkl_i - \br_t \Big\|_2^2,
\end{align}
subject to $\sum_{i=1}^{N} \eta_i = 1$,
where $\bwmkl_i$ is the weight vector corresponding to the $i$th feature
space (\ie\ kernel).  The justification for
using this algorithm is that we expect to use many kernels in the
first iteration rounds of our search and not too many near the end, as we gain a
better understanding of relevance inferred through (explicit) pointer
clicks and (implicit) eye movements (as described in
Section~\ref{section:relevance}).  After each iteration of the search, 
when the user has indicated the relevance of the newly seen images,
we can use this feedback as the labels (outputs) of our classification
(regression) MKL problem to find a new set of kernel weights $\etab$
(based solely on the images seen thus far). 

After we learn this new representation we supply these weighted
kernels to the kernelized \linrel~algorithm of
Section~\ref{s:LINRELAlgorithm}.  However, before that we describe the
component of our system that uses eye movements as an extra set of
features, by creating a combined space using the kernel learnt using
Equation (\ref{mkl}).

\subsection{Tensor Decomposition}
\label{section:tensor}

Since eye movements are available only for images
already presented to the user, eye movement features cannot be
used directly for predicting the relevance of unseen images.
To elevate this problem, we relate the known image features to the
(yet) unknown eye movement features by learning
a joint representation that combines
these two views.  
We
learn this relationship by using a tensor representation which creates
an implicit correlation space~\cite{Hardoon:2010b}. The tensor
representation can be computed by taking dot products between
each individual kernel matrix of each
view~\cite{Pulmannova,SS-JST-EPH_05}.

Hence, let $\bK_{\Phi_t} = [k^{\phi}(I_u,I_v)]_{1\leq u,v \leq t-1}$
be the kernel Gram matrix constructed from previously seen $t-1$ image
feature vectors
$\Phi_t = [\phi(I_\tau)]_{1\leq \tau \leq t-1}$.
Similarly, let 
$\bK_{\Psi_t} = [k^{\psi}(I_u,I_v)]_{1\leq u,v \leq t-1}$
be the kernel Gram matrix constructed from eye movement features
$\Psi_t = [\psi(I_\tau)]_{1\leq \tau \leq t-1}$.
Given these two kernel matrices we can combine them by taking a
component-wise product \linebreak \mbox{$\bK_{\Phi \circ \Psi} = \bK_{\Phi_t} \circ
\bK_{\Psi_t}$}, which corresponds to a \emph{tensor product} between
feature vectors $\Phi_t$ and $\Psi_t$.  We then use the kernel matrix
$\bK_{\Phi \circ \Psi}$ to train a tensor kernel
SVM~\cite{Hardoon:2010} to generate a weight matrix which is composed
of both views. As mentioned earlier, we do not have the eye movement
features for images not yet displayed to the user. Hence, we need to
decompose the weight matrix into one weight vector per view. This has
been resolved by~\cite{Hardoon:2010}, who propose a novel singular
value decomposition (SVD) like approach for decomposing the resulting
tensor weight matrix into its two component parts, without needing to
directly access the feature spaces $\phi$ and $\psi$.

Therefore, assume the weight matrix decomposed for the image features
is $\Alpha \in \real^{(t-1) \times D}$ and for the eye movement
features is $\Beta \in \real^{(t-1) \times D}$, where $t-1$
corresponds to the number of images seen and $D$ is the dimensionality
of the decomposition.  Given $\Beta = [\beta_u \in \real^D]_{1 \leq u
  \leq t-1}$ we can project any of the MKL combined image features as
follows~\cite{Hardoon:2010b}:
\[
\small \tilde{\phi}(I) = \left(\sum_{u=1}^{t-1} k_{\etab}^{\phi}(I_u,I) \beta^1_u, \ldots, \sum_{u=1}^{t-1} k_{\etab}^{\phi}(I_u,I) \beta^D_u \right),
\] 
to produce a new feature vector which has been mapped into the eye
movement feature space using the matrix $\Beta$.  Finally, we can
create the following kernel function $k_{\etab}^{\tilde{\phi}}(I,J) =
\tilde{\phi}(I)\tilde{\phi}(J)^\top$ from our new feature vectors
$\tilde{\phi}(I)$ and $\tilde{\phi}(J)$.  After we have this new
representation we can pass this new updated kernel to the kernelized
\linrel~algorithm of the following
section.

\subsection{The \linrel~Algorithm}
\label{s:LINRELAlgorithm}

After updating the similarity metric and the associated kernel through
MKL and the tensor decomposition as described in the previous
sections, the \linrel\ algorithm is used for selecting the next
collage of images that is presented to the user. The \linrel\
algorithm (originally devised and analysed in~\cite{auer:jmlr02}) is
an exploration-exploitation oriented online learning algorithm. It
aims to sequentially present images to the user such that the positive
feedback from the user is maximized. Hence the \linrel\ algorithm is
very well suited to be used in the PinView system for retrieving
images that are of interest to the user.

Given the image features, the relevance of images is assumed to be
mutually independent.
\linrel\ then assumes that the expected relevance~$r_I$ of an image~$I$ is given by an (unknown) linear function of the image features $\phi(I)$,
\[ \EXP{r_I} = \phi(I) \cdot \bw, \]
with an unknown weight vector~$\bw$. Thus in each step~$t$ of the search, \linrel\ estimates the weight vector by some $\hw_t$ and uses this estimate to select an image which is likely to be relevant. But since the estimate $\hw_t$ might be inaccurate, \linrel\ also needs to ensure a sufficient amount of exploration. This is achieved by taking into account a bound~$\sigma_t^2(I)$ on the variance of the estimated relevance~$\phi(I) \cdot \hw_t$, by considering an appropriate confidence interval for the ``true'' expected relevance~$\phi(I) \cdot \bw$. Thus \linrel\ selects the image which maximizes the upper confidence bound, 
\begin{equation} \label{eq:linrel-ucb}
I_t = \arg\max_I \left( \phi(I) \cdot \hw_t + c\sigma_t(I) \right), 
\end{equation}
where the parameter $c \geq 0$ controls the amount of exploration. This rule selects an image if its predicted relevance is high (which is an exploitative selection), or if the variance of this estimate is high (which is an explorative selection). It is shown in the analysis of \linrel~\cite{auer:jmlr02}, that selecting an image with high variance according to the above rule improves the accuracy of the estimated weight vector~$\hw_t$. It is also shown that the error rate of \linrel\ --- compared to the best linear predictor of the relevance --- is essentially~$\sqrt{d/t}$ after~$t$ steps of the search, where~$d$ is the number of dimensions of the feature vector~$\phi(I)$.
While the original \linrel\ algorithm in~\cite{auer:jmlr02} explores the dimensions of the feature vector explicitly, more recent variations of \linrel\ (e.g. LinUCB in~\cite{lcls-cbapnar-2010}) use regularization to deal with large feature vectors. For the PinView system we also use regularized \linrel, which calculates an estimate for the weight vector by regularized linear regression for the observed relevance scores of the selected images so far. The solution of the regularized regression can be written using the Gram matrix $\Phi_t \Phi_t^\top$ as 
\begin{equation}
 \hw_t = \Phi_t^\top (\Phi_t \Phi_t^\top + \mu \id)^{-1} \br_t,
 \label{eq:linrel-regression}
\end{equation}
where $\Phi_t$ is the matrix of feature vectors of the images selected so far, $\Phi_t = [\phi(I_\tau)]_{\tau=1,\ldots,t-1}$, $\br_t=(r_{I_1},\ldots,r_{I_{t-1}})$ is the vector of relevance scores observed so far, $\id$ denotes the identity matrix, and $\mu>0$ is the regularization parameter. Thus the estimated relevance of an image is given by 
\[ 
\phi(I) \cdot \hw_t 
= \phi(I) \Phi_t^\top (\Phi_t \Phi_t^\top + \mu \id)^{-1} \br_t 
= \ba_t(I) \cdot \br_t, 
\]
where 
\[ \ba_t(I) = \phi(I) \Phi_t^\top (\Phi_t \Phi_t^\top + \mu \id)^{-1}. \]
It has been shown in~\cite{auer:jmlr02} that the variance of the estimate~$\ba_t(I) \cdot \br_t$ can essentially be bounded by $\sigma_t^2(I)=\|\ba_t(I)\|_2^2$. Thus the selection rule~(\ref{eq:linrel-ucb}) of \linrel\ can be rewritten as 
\begin{equation} \label{eq:linrel-rule}
I_t = \arg\max_I \left( \ba_t(I) \cdot \br_t + c \|\ba_t\|_2 \right). 
\end{equation}
This rule can easily be kernelized to accommodate the kernels generated by MKL, since the Gram matrix $\Phi_t \Phi_t^\top$ can be expressed as the kernel matrix $[k^{\phi}(I_u,I_v)]_{1 \leq u,v \leq t-1}$ and $\phi(I) \Phi_t^\top = [k^{\phi}(I,I_u)]_{1 \leq u \leq t-1}$. 

Since in each iteration of a search the PinView system not only selects a single image but a collage of several images, the \linrel\ algorithm needs to be extended to accommodate this. An obvious extension --- implemented for the experiments reported in Section~\ref{section:experiments} --- is to select all images of the collage according to rule~(\ref{eq:linrel-rule}), while each image is selected at most once during the search. This method for selecting a collage rather emphasizes exploration, since all images of the collage are selected by taking also an exploration term into account. An alternative method would be to select only one image according to rule~(\ref{eq:linrel-rule}), and to select the remaining images to maximize the estimated relevance $\ba_t(I) \cdot \br_t$. This second method selects at most one explorative image, and is thus far less exploratory then the first method. By selecting more than one image according rule~(\ref{eq:linrel-rule}), it is possible to interpolate between the first and the second method. Future work will show which collage selection method is most beneficial.

\section{Experiments} 
\label{section:experiments}

In this section we describe experimental evaluations of the PinView
system. We study empirically the following two questions:
\begin{enumerate}
\item[(a)] How close to explicit feedback performance can we get with
  less laborious implicit feedback?
\item[(b)] Is it possible to still improve performance by combining
  implicit and explicit feedback, especially when the explicit
  feedback is only partial (a single click on the most relevant
  image) and gaze patterns can be expected to reveal more relevant
  images.
\end{enumerate}

In the experiments we use three variants of PinView:
\begin{enumerate}
\item PinView system with implicit feedback from gaze patterns.
\item PinView system with explicit feedback from clicks.
\item PinView system with both explicit and implicit feedback, from
  both gaze patterns and clicks.
\end{enumerate}
For evaluation purposes these variants are compared with the baseline
of browsing (that is, showing randomly ordered images) and the
PicSOM~\cite{laaksonen:networks02} CBIR system sharing the same
interface as PinView but lacking the novel machine learning
components. This way, the comparison emphasizes the effects caused
by the new components instead of the interface.

To keep the experimental cost
manageable, we started with extensive offline experiments and then
validated the main findings later in online experiments with real
users -- performing all the comparisons with online users would not have
been feasible.  In offline setups we choose relevance of
images based on their tags or classes, and simulate the feedback based
on the relevance. Explicit feedback comes directly from the relevance
and for implicit feedback we use eye movement features computed from
relevant and nonrelevant images viewed in earlier experiments. We
expect the simulated explicit feedback to be a reasonable
approximation to real feedback, and hence in the online experiments we
focus on validating the implicit feedback results.

\subsection{Offline Experiments}

\textbf{The data set} of images used in the offline experiments is the
\emph{train} subset of the PASCAL Visual Object Classes Challenge 2007
(VOC2007) dataset~\cite{pascal-voc-2007}.  The number of images in
this dataset is 2501.  It contains 20 overlapping categories whose
summary statistics are given in
Table~\ref{table:vocdatasetstatistics}.

\begin{table}[t]
\caption{Summary of the subset of the VOC2007 dataset used in offline experiments.}
\label{table:vocdatasetstatistics}
\begin{center}
\begin{tabular}{lcc}
\hline
Category name & Number of images & Percentage of images\\
\hline
Cat & 166 & 6.6 \\
Dog & 210 & 8.4  \\
Cow & 71 & 2.8  \\
Horse & 144 & 5.8 \\
Person & 1070 & 43.1 \\
Bird & 182 & 7.3 \\
Sheep & 49 & 2.0 \\
Aeroplane & 113 & 4.5 \\
Bicycle & 122 & 4.9 \\
Boat & 87 & 3.5 \\
Bus & 100 & 4.0 \\
Car & 402 & 16.1 \\
Motorbike & 123 & 5.0 \\
Train & 128 & 5.1 \\
Bottle & 153 & 6.1 \\
Chair & 282 & 11.3 \\
Diningtable & 130 & 5.2 \\
Pottedplant & 153 & 6.1 \\
Sofa & 188 & 7.5 \\
Tv-monitor & 144 & 5.8 \vspace{1mm}\\
\hline
All   & 2501 & 100.0 \vspace{1mm}
\end{tabular}
\end{center}
\vskip -0.1in
\end{table}

\textbf{Experiment setup:} Each offline experiment consists of
simulated search sessions.  In each search session PinView selects ten
collages with 15 images each.  
The goal of a search session is to retrieve images from one of the
categories. For simulating user feedback, images are divided into
relevant images (all those from the desired category) and non-relevant
images (all those not from this category). The calculation of
different feedback modalities is detailed below.
In each experiment the performance of the retrieval systems is
measured in 40 search sessions on each of the 20 categories.

The regularization parameters of MKL and \linrel{} are set to a
single combined regularization parameter which is found for each
feedback modality with a grid search over values
$\{0,0.01,0.1,5,10,100,1000\}$.

\textbf{Feedback modalities:} The following versions of PinView were compared:
\begin{enumerate}
\item Implicit feedback from simulated eye movements: {\sc
    SimulatedEye}.  The simulated eye movements are selected from a
  pool of previously recorded eye movements from online experiments.
  The eye movements are split to two groups, ``positive'' and
  ``negative'', depending on whether the viewed image was relevant or
  nonrelevant in the task in which it was recorded.  Both of these
  groups are divided into five subgroups depending on how many
  relevant images there were in the collage where the image was seen;
  the rationale is that the eye movements differ between collages
  having significantly different numbers of relevant images.  The
  subgroups correspond to the following number of relevant images on a
  collage: 0, 1, 2--3, 4--6, 7--10, and 10--15.  In the experiment,
  eye movements are sampled from the positive group for relevant
  images and from the negative group for nonrelevant ones, taking into
  account the number of relevant images in the current collage.

\item Explicit feedback from simulated clicks based on the known
  relevances of the images: {\sc SimulatedClick}. To simulate an
  interface that still retains a low level of manual effort the system
  operates in a mode where only one of the images is clicked. If there
  are several relevant images, a random relevant image on the collage
  is selected as clicked.  If the collage contains no relevant images,
  then an image is picked uniformly at random.

\item Combined explicit and implicit feedback: {\sc
    SimulatedEye+Click}. Here both types of input are simulated, and
  used in the model as in Eq.~(\ref{eq:eye-prediction}).  The explicit
  click weight $\alpha$ of the model is found by running a
  grid search over the values $\{0.01,0.1,1,5,10,100\}$, before
  choosing the regularization parameter of the PinView system.

\item For completeness we additionally include one more type of
  explicit feedback: {\sc Full}, where the true class label of each
  seen image is given, corresponding to explicit feedback in which
  each relevant image is clicked.
\end{enumerate}

Of the simulated feedbacks, {\sc Full} feedback simulates the
performance of PinView under ideal conditions, where the user is able
and willing to provide perfect feedback.  The other simulations
provide lower bounds for the performance obtainable using only the
implicit feedback, or by the partial explicit feedback of a
single click that is still relatively effortless to provide.  The real
performance of the system in online experiments is expected to lie
between these two extremes.  This is because the simulated runs use
only the incomplete tag information; in a real system the user is also
able to give more refined feedback due to his ability to use the
visual content of the images. Given a collage with more than one
relevant image the user will not make the choice randomly, but will
base his decision on the content, and the eye movements will also
reflect the relative similarity of the images and the search target
not captured by the simulation process.

\textbf{Evaluation.} To evaluate the model we record the performance
on each feedback type separately.  The measure of performance is mean
average precision (MAP), i.e., the average fraction of relevant images
that the system returned, averaged over the found relevant images and
search sessions.  

\begin{figure*}
\centering
\includegraphics[width=0.75\textwidth]{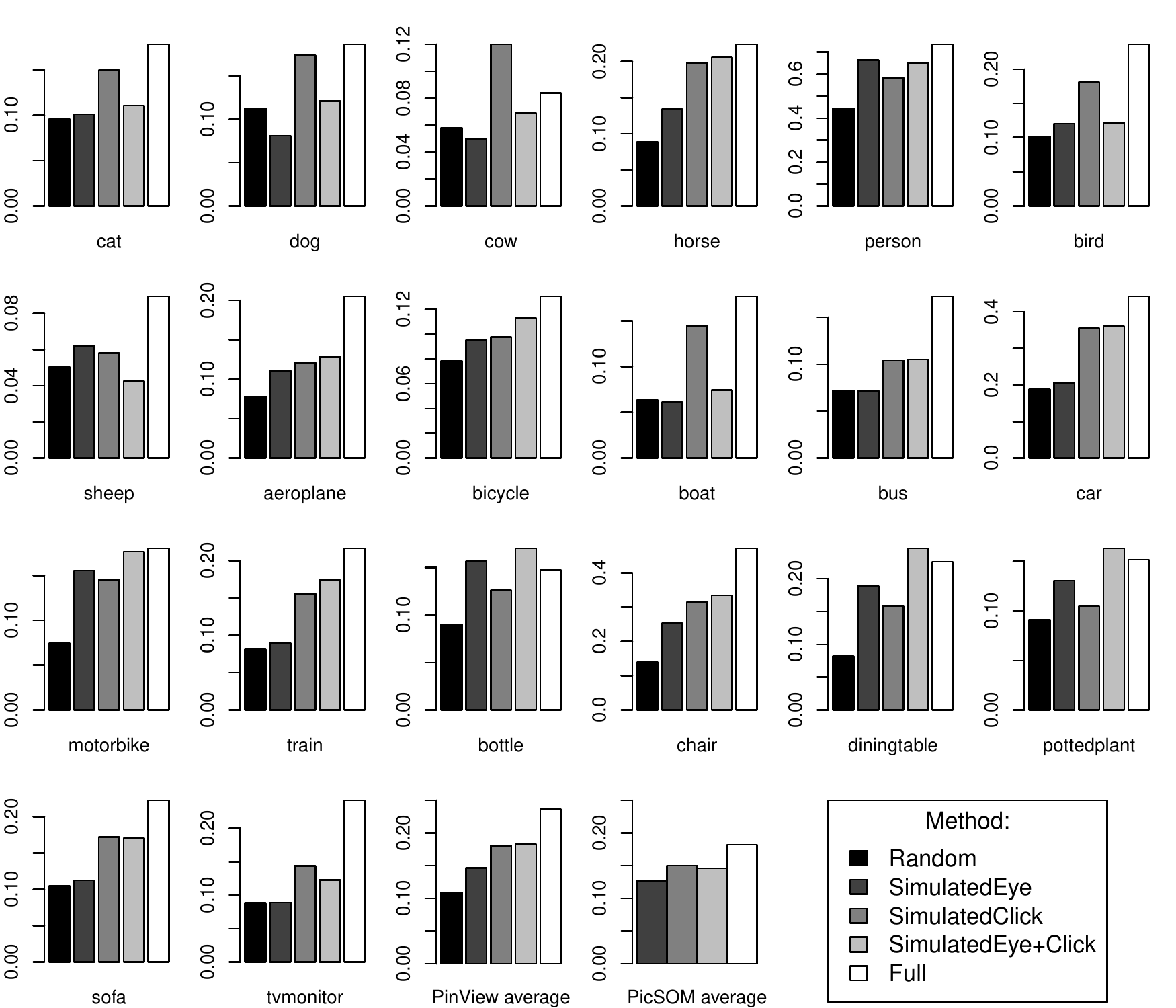}
\caption{Mean averaged precision (MAP) of the PinView variants during offline
  experiments in each of the search tasks, compared to pure browsing
  ({\sc Random}) and {\sc Full} feedback for each image. The last two
charts collect the macro-averaged performance of the PinView system and the PicSOM
comparison method. 
}
\label{figure:offlineresults}
\end{figure*}

\begin{table}
\caption{Pairwise paired t-test p-values for methods in the offline
  experiment.  The empirical performance of the methods increases from
  left to right and from top to bottom, and SimulatedEye is abbreaviated
as SimEye to save space.}
\label{table:ttests:offline}
\begin{center}
\begin{tabular}{|l|ccc|}
\hline
 & SimEye & Click & SimEye+Click\\
\hline
Random & 0.0085 & 1.5e-06 & 8.2e-05 \\
SimEye & \textbullet & 0.011 & 6.6e-04   \\
Click &  \textbullet & \textbullet & 0.80 \\
\hline
\end{tabular}
\end{center}
\vskip -0.1in
\end{table}

The \textbf{results of the offline experiments} are presented in
Figure~\ref{figure:offlineresults}. As expected, all PinView results
lie in between the (laborious) {\sc Full} feedback results and pure
browsing results ({\sc Random}). Implicit feedback ({\sc
  SimulatedEye}) outperforms browsing, even if it does not provide as
good feedback as pure explicit feedback ({\sc SimulatedClick}). These
differences are significant
(Table~\ref{table:ttests:offline}). Combined explicit and implicit
feedback ({\sc SimulatedEye+Click}) gives very similar results to pure
explicit feedback, and the small difference between the two is not
significant.
All of the reported results were run without the tensor decomposition
of Section~\ref{section:tensor}, since it did not increase the overall
performance of the system despite showing improved accuracy for some
users and tasks.

Comparing the MAP results between the PinView and PicSOM algorithms
(reported as averages over all tasks in Figure~\ref{figure:offlineresults}),
it is evident that the here-proposed PinView algorithm is superior
with all the feedback modalities.
Most importantly, PinView seems to be better than PicSOM in making
simultaneous use of both explicit and implicit relevance feedback,
which can be seen when comparing the \textsc{SimulatedClick} and
\textsc{SimulatedEye+Click} results.

\subsection{Online Experiments}

In this section we describe online experiments in which test subjects
interact with the PinView system. The goal of the online experiments
is to validate the offline findings about relative goodness of the
different feedback modalities, and naturally also to give evidence of
how well the system works in practice.

\textbf{Data sets.}  The online experiments use a subset of the
ImageNet dataset~\cite{deng:imagenet09}, created by the authors and
called IMG2010 dataset. It contains 3720 images from several
categories (synsets) of the ImageNet, which is a database containing
URLs to images available on the internet together with semantic
category information (synsets of WordNet) and a hierarchy between the
categories.  Hence, IMG2010 contains images that are representative of
ones that appear on the internet.

\textbf{Experiment setup and evaluation.}  Each of the ten users
performed 12 different search tasks which mimic different real-world
scenarios. The tasks ranged from scenarios where a tag-based search had
first been used to prune the eligible images, to scenarios where the
images were more diverse.  During one search task the system showed to
the user a total of 120 images, contained in eight separate collages
each having 15 images.  Before the search session the system instructed
the user to find a shown target image that belongs to a given
category.  In practice, the experiment took approximately 20--30
minutes per subject.  The 12 search tasks were divided into four
groups, each consisting of three sub-tasks.  The four groups were:
\begin{enumerate}
\item Finding images of a particular sport from among sports
  images. The particular sport categories were ice hockey, gymnastics,
  and soccer. The image dataset for this group contains 1006 images
  sampled from the sports subcategory of ImageNet, which has 89 ice
  hockey, 92 gymnastics, and 88 soccer images.
\item Finding images of aeroplanes.  The image dataset contains 900
  uniformly sampled images that are not flowers or aircraft, and
  additional 150 images of both aircraft and flowers.
\item Finding images of flowers.  The image dataset is the same as in
  the previous group.
\item Finding images of a given mammal, amongst other mammal
  images. The goal categories are deer and cheetah (twice).  The
  dataset contains 105 images sampled from deer category, 99 images
  sampled from cheetah category, and 612 images sampled from a mammal
  category that are not deers or cheetahs.
\end{enumerate}

As the goodness criterion we again used the number of relevant images.
The different PinView variants were randomly allocated to the sub-tasks
so that each sub-task had as uniform allocation of variants as
possible.  The regularization parameter was set for each PinView
variant to the value that performed the best in the offline
experiments.

\begin{figure*}
\centering
\includegraphics[width=0.75\textwidth]{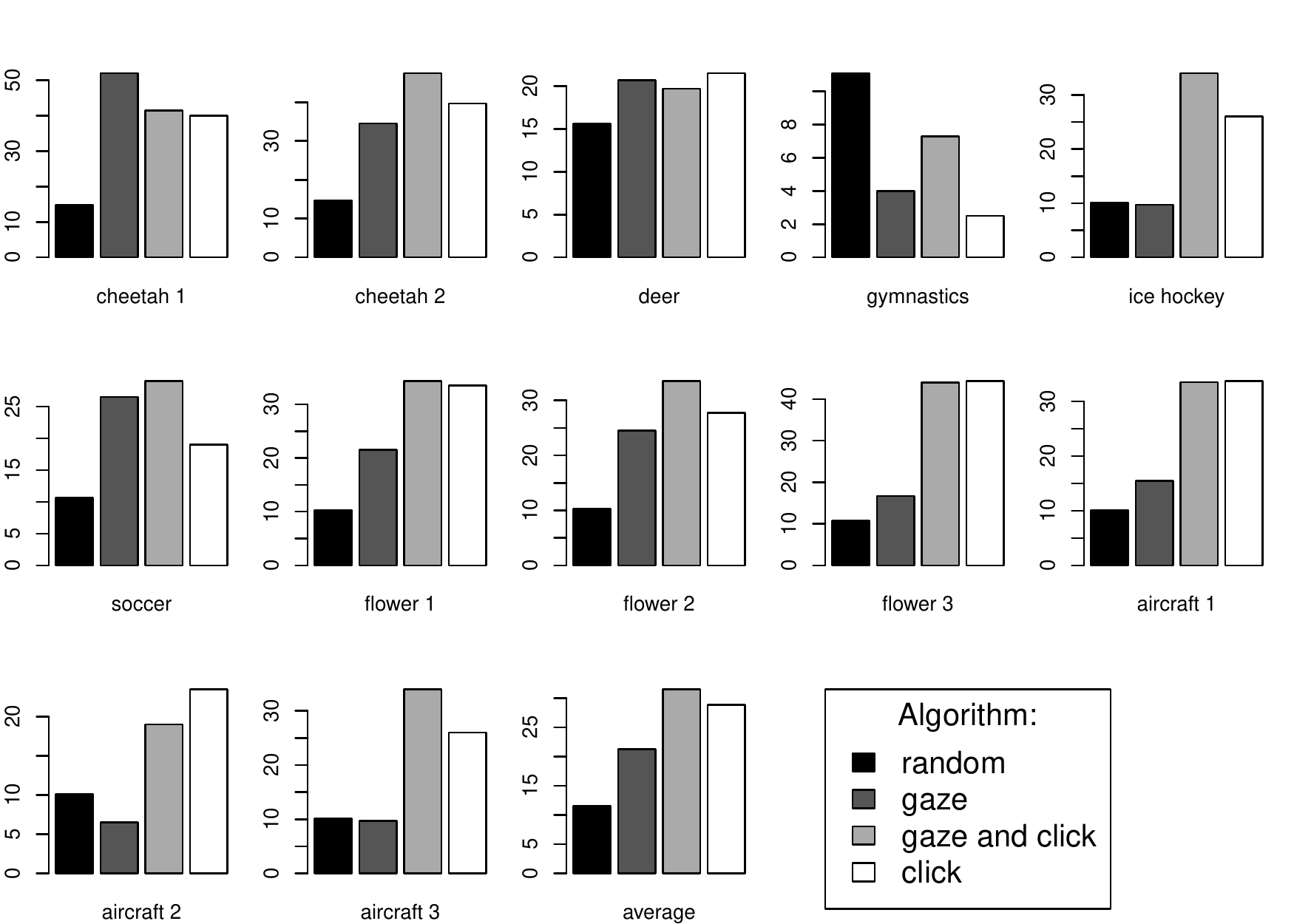}
\caption{Number of retrieved relevant images in each task in online
  experiments. Each plot corresponds to a task and each bar
  corresponds to a PinView variant or the baseline of browsing
  randomly ordered images. The total number of retrieved images in
  each task is 120.
}
\label{figure:onlineresults}
\end{figure*}

\textbf{Results.}  The quantitative performance of the PinView
variants is shown in Figure~\ref{figure:onlineresults} for each task.
All input modalities used by PinView are clearly better than the
baseline of browsing randomly ordered images, which is confirmed by
t-tests in Table~\ref{table:ttests}.  Only in one of the tasks,
gymnastics, the performance of PinView was below the baseline, which
might be due to random fluctuations because of noise. 

The relative performance of the variants varies between tasks.
Implicit feedback from gaze is worse than explicit feedback from
clicks, although the difference is not strongly significant (t-test,
$p=0.046$). The number of relevant images retrieved by gaze is on
average 67\% of the number of relevant images returned by the best
modality (the combined click and gaze).  However, the gaze feedback
performs well in many of the tasks and hence gaze is a viable source
of implicit feedback information.  

The paired t-test gives a p-value of $0.059$ on the hypothesis that
the performances of the click modality and combined click and gaze
modality are the same.  There is evidence that combining information
from click and gaze modalities improves the performance of the system,
but more extensive testing would be needed for strong conclusions.
The performance of the combined click and gaze modality is relatively
better in online than in offline experiments, which might be due to the
fact that the relevance feedback given by real users is more accurate
than the simulated one.

\begin{table}
\caption{Pairwise t-test p-values for methods in the online
  experiment.  The empirical performance of the methods increases from
  left to right and from top to bottom.  }
\label{table:ttests}
\begin{center}
\begin{tabular}{|l|ccc|}
\hline
 & Gaze & Click & Gaze and click\\
\hline
Random & 0.016 & 2.6e-04 & 7.5e-05 \\
Gaze & \textbullet &  0.046 & 0.0065  \\
Click &  \textbullet & \textbullet & 0.059   \\
\hline
\end{tabular}
\end{center}
\vskip -0.1in
\end{table}

\section{Discussion and conclusions}
\label{section:conclusion}

In this paper we described our PinView CBIR system
which records implicit relevance signals from the user and
infers his image search intent by using several novel machine learning
methods.  We show that the PinView variants work better
than browsing (a set of randomly ordered images), indicating that
PinView would be useful at least in scenarios where tag-based evidence
is not available or has already been used to narrow down the search to
a subset of the original collection. 

Implicit feedback from gaze outperformed the baseline,
suggesting that pure implicit feedback is a viable option when
it is difficult or too laborious to give explicit feedback. Explicit
feedback by clicks gave more accurate results, and there was evidence
that combined explicit and implicit feedback produced the best
results. In summary, the compilation of algorithms in PinView is a very
promising approach to content-based image retrieval.
One of the main use scenarios is a search session where first a
tag-based search is used to focus on a subset of potentially relevant
images, and content-based search is then needed to do further
retrieval in the still large result set. 

Our final conclusions from the present work and other serious
attempts~\cite{DBLP:phd/de/Essig2007,OyekoyaPhD07} to use and evaluate
implicit relevance feedback from eye movements in iterative online
content-based image retrieval are as follows:
First, when used for purely implicit relevance feedback, eye movements
perform better than random picking as was demonstrated
in~\cite{OyekoyaPhD07} and in this paper.  This mode of operation can
prove to be useful if the setup does not allow giving explicit
feedback, or if the relevance feedback mechanism is used to secretly
improve the efficiency of otherwise random browsing.
Second, the performance level of gaze-based implicit relevance
feedback with current hardware and algorithmic techniques cannot reach
that of click-based explicit feedback.

Third, when combining explicit click-based and implicit gaze-based
relevance feedback together, the system performance will exceed the
level of solely explicit relevance feedback as was proven
in~\cite{DBLP:phd/de/Essig2007} and in our experiments.  To what extent this
happens most likely depends on the experiment arrangements, including
the data set, eye-tracking device, and the user interface design.
In~\cite{DBLP:phd/de/Essig2007}, the image collection was arguably
simpler than ours.  Additionally, the user interface allowed the use
of gaze for comparing the query image and the candidates, which surely
was beneficial for the proposed method.
We thus argue that our experiments have resembled genuine use
scenarios of content-based image retrieval, with respect to both the
used data and the user interface, more than the previous attempts.
We also argue that we have been able to show that even in such a
difficult context, gaze tracking data has proven to be a useful source
of implicit relevance feedback that can be beneficially used either
alone or together with explicit feedback.

{
\section*{Acknowledgments}
The research leading to these results has received funding from the
European Community's Seventh Framework Programme (FP7/2007--2013)
under \emph{grant agreement} n${}^\circ$ 216529, Personal Information
Navigator Adapting Through Viewing, PinView, IST Programme of the
European Community, under the PASCAL2 Network of Excellence,
IST-2007-216886, and the Academy of Finland for the Finnish Centre of
Excellence in Computational Inference Research (COIN, 251170).  This
publication only reflects the authors' views.  }


\end{document}